\documentclass[preprint2]{aastex}

\slugcomment{Version 3.3  2007/05/03}

\shorttitle{GLIMPSE Identification of SiO Maser and OH/IR Objects}
\shortauthors{Deguchi et al.}

\begin{document}


\title{Identification of Very Red Counterparts of SiO Maser and OH/IR Objects in the GLIMPSE Survey}

\author{Shuji \textsc{Deguchi}%
}
\affil{Nobeyama Radio Observatory, National Astronomical Observatory \\
Minamimaki, Minamisaku, Nagano 384-1305, Japan}
\author{Jun-ichi \textsc{Nakashima} }
\affil{Academia Sinica, Institute of Astronomy
and Astrophysics, PO Box 23-141, Taipei, 106, Taiwan}
\author{ Sun {\sc Kwok}}
\affil{Department of Physics, University of Hong Kong, Hong Kong, China\\ and \\
Department of Physics \& Astronomy, University of Calgary, 
Calgary, Alberta, Canada T2N 1N4}

\author{Nicolas \textsc{Koning}}
\affil{Department of Physics \& Astronomy, University of Calgary, 
Calgary, Alberta, Canada T2N 1N4}
%
\author{ }
\affil{(ApJ part 1,  August 10, 2007, v665 no.1 issue in press)} 



\begin{abstract}
Using the 3.6/4.5/5.8/8.0 $\mu$m images with 1.2\arcsec\ pixel resolution 
from the Spitzer/GLIMPSE survey,
we investigated 23 masing and 18 very red objects
that were not identified in the 2MASS survey.
Counterparts for all selected objects were found in the GLIMPSE images.
Color indices in these IR bands suggest the presence of a high-extinction layer 
of more than a few tenths of a solar mass in front of the central star. 
Furthermore, radio observations in the SiO and H$_2$O maser lines found
characteristic maser-line spectra of the embedded objects, e.g., the SiO $J=1$--0 line intensity 
in the $v=2$ state stronger than that of the $v=1$ state, or very widespread H$_2$O maser emission spectra.
This indicates that these objects are actually enshrouded by very thick circumstellar matter,
some of which cannot be ascribed to the AGB wind of the central star. 
Individually interesting objects are discussed, including two newly found water fountains 
and an SiO source with nebulosity.
\end{abstract}





\section{Introduction}

The discovery of stellar maser sources \citep{wil68} and their association 
with very red infrared objects has led to a new class of objects called OH/IR stars.  
From the OH profiles, we came to the realization that the maser emissions 
are a manifestation of the mass loss process \citep{kwo76}, 
and OH/IR stars are likely to be stars in the asymptotic giant branch (AGB) phase 
of evolution \citep{hab96,sev02}.  The mass loss process also 
creates a thick circumstellar dust envelope which often obscures the central star.  
Although OH/IR stars are often bright mid-infrared objects \citep{jim06}, 
many of them have no optical or near infrared counterparts.  
In the absence of a spectral classification based on photospheric spectroscopy, 
the nature of these stars is classified based on the circumstellar spectra 
in the mid-infrared \citep{kwo97}.

It should be noted that the mid-infrared colors alone are  not enough to determine the nature of maser sources.  
For example, the IRAS colors of mass-losing evolved stars overlap 
with colors of young stellar objects embedded in molecular clouds \citep{van88}, 
and there is no well-defined boundaries in the color-color diagram between young 
and evolved stars \citep{wei90}.  The maser characteristics by themselves are not enough 
to distinguish the two classes of objects \citep{cas99}.   
The situation for SiO maser sources is also complex.  
For example, there are three SiO maser sources in molecular clouds \citep{has85,mor92} 
that are likely to be young stellar objects.  There are also peculiar cases, such as the SiO maser source IRAS 19312+1950 is 
an evolved star in a dark cloud \citep{nak04,deg04c}, and 
V838 Mon is an M supergiant with SiO emission which emerged after nova eruption \citep{deg05b,tyl06}.

In some cases, the imaging and identification of the central star in NIR bands can greatly facilitate 
the determination of  the nature of these objects \citep{lew04,deg05a}. 
Young, low-mass stars are usually born in a star cluster, but evolved stars, 
especially AGB/post-AGB stars with the ages of Giga years, are observed 
as a single object \citep[see $\S$ 3 of ][]{deg04c}.   
However, for the case of supergiants, they are frequently a member 
of stellar associations \citep{hum70}, suggesting they are 
on their way towards leaving the originating star clusters. Deep infrared imaging 
can help find the surrounding low-mass stars.

The recently performed Galactic Legacy Infrared Mid-Plane Survey Extraordinaire (GLIMPSE) 
of the {\it Spitzer Space Telescope (SST)} has provided unprecedented deep near-infrared 
images of the Galactic plane.  Because of the high angular resolution 1.2\arcsec\ per pixel) 
of the GLIMPSE observations, we have the opportunity to obtain an accurate identification 
of the infrared/maser sources.  In this paper, we report the identification of 41 infrared sources from the GLIMPSE data, as well as SiO or H$_2$O maser observations of these sources with the Nobeyama 45-m telescope.

\section{Observations}
\subsection{Source selection}
We selected 41 sources from the {\it Midcourse Space Experiment (MSX)} Galactic plane survey that are found to have no (or dubious) near-infrared counterparts in the 2MASS sky survey \citep{skr00}. 
Half of these (22 objects) are OH/IR objects listed by \citet{sev01}, \citet{tel91}, \citet{ede88}, etc. 
The positional accuracy of the MSX positions is about a few arcseconds \citep{ega99}, comparable
to the positional accuracies of OH masers measured with the Very Large Array (VLA) [see figure 9 of \cite{deg02} for comparison].  
Most of these objects have reasonably red mid-IR colors with 
$C_{CE} \ [\equiv log(F_E/F_C)]$ between 0 and  0.5,  where $F_C$ and   $F_E$
are the MSX band C (12 $\mu$m) and band E  (21 $\mu$m) flux densities, respectively.

In addition to the MSX sources, we have included a number of medium-bright and red IRAS sources without 2MASS counterparts for which no maser line search was made.
These are IRAS 18030$-$1858,
18151$-$1713, 18241$-$1010, 18298$-$0904, 18424$-$0509, 18444$-$0359, 19011+0638,
19025+0702, 19087+1006, 19112+1220,
19114+0920, 19180+1230, 19254+1724, 19266+1815, 19275+1720, G014.2758$-$00.1602,
G027.8433$-$00.1060.  The MSX colors of these objects are not much different from those of the above OH/IR sources 
without NIR counterparts.
The entire sample we chose for identification in the Spitzer/GLIMPSE survey is listed in table 1,
giving the source name, MSX(6C) name, right ascension and declination (J2000) of the MSX source, flux density in the MSX C (12$\mu$m) band,
the MSX colors [$C_{AC}=log(F_C/F_A)$, and $C_{CE}=log(F_E/F_C)$], status of detections in SiO, H$_2$O, and OH masers,
and a conventional OH object name.

\subsection{Spitzer/GLIMPSE Identification}
The GLIMPSE survey was made with SST between March and November, 2004 using the Infrared Array Camera (IRAC) \citep{ind04}. 
The survey covered  the area of $|l|=10$--65$^{\circ}$ over latitudes $|b|<1^{\circ}$
toward the inner Galactic disk excluding the Galactic center. Simultaneous $5.2' \times 5.2'$ images at 3.6, 4.5, 5.8 and 8 $\mu$m were obtained at a spatial resolution of  $\sim 1.2'' \times 1.2''$ per pixel. 

The color-composite images (from the 3.6, 4.5 and 5.8 $\mu$m band images)  
toward the 41 selected objects are shown in Figure 1. The counterparts for all sources in our sample
were found; they are all very red and near the center of the images.
Because all of these objects are bright in 8 $\mu$m, the identifications
in other bands were made quite easily. The color images were created
with special software developed at the Space Astronomy Laboratory of the University of Calgary. 
The flux densities were derived from the GLIMPSE archives;
the aperture size used for photometry was typically a few to 12 arcseconds, depending on the image size
of the star.  The estimated error of the photometry is dependent on the background noise levels, 
and are typically $< 2\%$ for the objects brighter than 1, 3, 7, 16 mJy at the 3.6, 4.5 and 5.8 $\mu$m bands respectively.   
The flux densities are given in columns 4--7 in table 2.  
The color indices, $[3.6]-[5.8]$ 
and $[3.6]-[12]$, are also listed in the 
8th and 9th columns in table 2. Here, [3.6], [5.8], and [12] are the magnitudes 
in the GLIMPSE 3.6 and and 5.8 $\mu$m, and MSX 12 $\mu$m bands, respectively,
and the flux densities at 0th magnitude are 280.9, 115.6, and 26.4 Jy in these bands, respectively.
The interstellar extinction toward each object
is given in the 10th column of table 2 and  
it is estimated by the formula 
based on the full-sky 100 $\mu$m composite map
of COBE/DIRBE and IRAS maps \citep{sch98}.\footnote{
available at {\it http://nedwww.ipac.caltech.edu/forms/calculator.html}.
}     

After getting accurate GLIMPSE positions (the 2nd and 3rd columns in table 2), 
we again checked the 2MASS images and 
found the red counterparts corresponding to the positions of 4 sources; 18286$-$0959 ($J$18312292$-$0957194),
18298$-$0904 ($J$18323700$-$0902301), 19275+1902 ($J$19294645+1709013), and 19266+1815 ($J$19285303 +1821228).
These counterparts are very faint in the K-band ($K=12.6$--14.7), 
making previous identifications uncertain. Their positions 
coincide with those of the GLIMPSE objects within 0.5\arcsec. 
\citet{jim06} listed NIR counterparts for 5 objects in the present sample:
18182$-$1504, 18460$-$0254, 18488$-$0107, 19087+1006, and 19254+1631.
Because of the faintness of the objects ($K>13$), their identifications were again doubtful. 
We have checked the GLIMPSE positions of these objects and found that 18182$-$1504, 18460$-$0254,
and 19254+1631 were correctly identified, but 18488$-$0107 and 19087+1006 were misidentified;
the GLIMPSE objects are located 5\arcsec NE and 6\arcsec S of the 2MASS stars ($K=$13.05 and 14.10) given by \citet{jim06}.

In addition to the above selected objects without 2MASS counterparts, we checked the GLIMPSE images of 
about 200 SiO maser sources within the area of $l=10$ -- 60$^{\circ}$, and $|b|<1^{\circ}$ where 
the near-infrared identifications are already available \citep{deg98,deg01,deg02}. The GLIMPSE images 
mostly exhibit a single bright object at the center of the field, except for IRAS 19027+0517,
which shows accompanying nebulosity (Fig.~6).  Because this object has a NIR counterpart
in 2MASS images and seems to be intrinsically different from the objects in our sample,
we gave a discussion of this object separately in Appendix A.  

We wish to make a special reference to IRAS 18450$-$0148 (W43A), which is known as a water maser fountain
with collimated outflows \citep{ima02}. The detection of SiO maser emission in this object
\citep{nak03b} suggests that this is an evolved star.
The near-infrared counterpart of this interesting object is identified here for the first time.

\subsection{SiO and H$_2$O maser observations at Nobeyama}
Because a majority of objects in our sample have not been searched for
SiO or H$_2$O maser emissions before, we observed the objects in 
SiO $J=1$--0 $v=1$ and 2 and H$_2$O $6_{16}$--$5_{23}$ maser lines with the 45-m  telescope 
at Nobeyama during April 10--23, 2006.  A cooled HEMT receiver (H40) 
and an accousto-optical spectrometer array were used for the 43 GHz observations.  
The receiver system temperature was about 200 K and the effective velocity resolution is 0.3 km s$^{-1}$.
The half-power beam width (HPBW) at 43 GHz was $\sim$40$''$ and the conversion factor 
of the antenna temperature to the flux density was $\sim$2.9 Jy K$^{-1}$. 
In addition to the SiO maser observations, we made a 22.235 GHz
H$_2$O maser observation on April 20 and 23, when weather was unfavorable for 43 GHz observations.
We used a cooled HEMT receiver (H22) and the same accousto-optical 
spectrometer array (with an effective velocity resolution of 0.6 km s$^{-1}$). 
The conversion factor of the antenna temperature to the flux density was 
$\sim$2.8 Jy K$^{-1}$ at 22 GHz.  Because  the H40 and H22 receivers have a 2 GHz band width,
we configured the AOS-H spectrometer array to have the spectra of the SiO $J=1$--0 $v=0$ (43.423 GHz), 
and $v=3$ (42.519 GHz), $^{29}$SiO $J=1$--0, v=0 (42.880 GHz), 
and H53$\alpha$ (42.952 GHz) lines simultaneously in the H40 receiver,
and NH$_3$ \ $J_{K}=1_1$--$1_1$, $2_2$--$2_2$, and $3_3$--$3_3$ lines 
(23.694, 23.722, 23.870 GHz, respectively) simultaneously in the H22 receiver.  
The rms noise temperatures for these additional lines are similar to the noise levels of the SiO and H$_2$O  lines 
and therefore are not repeated in tables 3 and 4.
No detections in these additional lines were obtained except for 18182$-$1504.  
Further details of observations using the NRO 45-m telescope are
described elsewhere \citep[e.g., ][]{nak03a}.

Among the 31 objects observed, 8 were detected in SiO and their spectra are shown in Fig. 2.
In all cases, the $v=2$ maser line is stronger than the $v=1$ maser line.
Water masers in the $6_{16}$--$5_{23}$ transition were detected in two sources 
(18286$-$0959 and 18460$-$0151) and their spectra are shown in Fig. 3.
It is notable that both of the detected objects exhibit a wide velocity range 
in the H$_2$O maser emission spectra, 
which resemble the H$_2$O spectra of  ``water fountains'' \citep{mor03}.   

Because some of our sample objects might be associated with young stellar objects
in molecular clouds, we also made simultaneous observations of the HCO$^+ \; J=1$--0 line 
at 89.189 GHz and the SiO $J=2$--1 $v=1$ line.
The non-detection of the HCO$^+$ line effectively excludes the possibility 
of these water maser sources being associated with young stellar objects.

\section{Discussion}

\subsection{Two-color diagram}
The selected objects have extremely thick circumstellar envelopes.
In addition, they suffer from a large interstellar reddening
in the Galactic plane even at NIR to MIR wavelengths. Therefore, it is hard to separate the
circumstellar from the interstellar reddening for these objects because of the uncertainty of the 
interstellar extinction correction.\footnote{
As noted in Section 2.2, we used a formula given by \citet{sch98}, which gives almost the maximum
value of interstellar extinction due to thick dust clouds toward each source. 
}
 We discuss here quantities of
the sampled objects by introducing the color indices, $[3.6]-[5.6]$ and $[3.6]-[12]$,
which vary relatively mildly with the large extinction.
We apply a wavelength-dependent extinction derived from the GLIMPSE survey
which was obtained from the color excess of stars at $l=42^{\circ}$ and 284$^{\circ}$
\citep{ind05}. 
We compare physical quantities of the sample objects  
with those of the proto-typical OH/IR star with SiO masers,
OH 127.8+0.0, which is believed to be at the final stage of evolution on the AGB.  
Physical parameters of this star are relatively well known \citep{suh02}, with
the optical depth of the envelope being about 10--15 at 9.8 $\mu$m. 
This bright OH/IR star ($F_C=184$ Jy) has $[3.6]-[5.8]=3.46$ and $[3.6]-[12]=5.0$.
Because of the position ($l\sim 128^{\circ}$), the interstellar extinction 
at the NIR wavelengths is small for this object ($A_K=0.64$).
If this object were behind a dust cloud 
with a heavier extinction of $A_K >7.6$, it would be undetectable on the 2MASS image ($K>13.7$).
The spherically symmetric model for OH 127.8+0.0 \citep{kem02} gave a mass loss rate
of $7 \times 10^{-5}$ M$_{\odot}$  yr$^{-1}$. The same model 
gives a total mass of the envelope of 0.2 M$_{\odot}$
(if integrated to the outer radius of the dust shell, $1.3\times 10^{17}$ cm,
as used by \citet{kem02}).  

Figure 4 shows a two-color diagram, $[3.6]-[12]$
versus $[3.6]-[5.8]$, for the sampled objects, where
OH 128.7+0.0 occupies the position indicated by the square
in this diagram. The interstellar reddening moves the position of the star 
to the upper right with an inclination of about unity because the extinctions
at 5.8 and 12 $\mu$m are comparable \citep[Table 1 of ][]{mat90}.  
Selected objects in this diagram distribute from the lower-left to the upper right with a
steeper slope ($\sim 1.71 \pm 0.19$) than the slope of the interstellar reddening line ($\lesssim 1$).
They appear in a very wide range of color indices, over 3 in $[3.6]-[5.8]$
(corresponding to $A_K=20$), which is too large to be considered an effect 
of interstellar reddening. These facts suggest that the extremely red colors of these objects 
cannot be explained solely by the interstellar reddening applied to the OH 127.8+0.0 type star, 
but they are more or less intrinsic to these objects. 

The steepness of the distribution seems to be understandable by the model that the 12 $\mu$m flux is
a re-emission of absorbed NIR light by dust grains near the central star.
If MIR re-emission does not occur (as interstellar reddening), the star should move
on the line indicated by the dotted arrow.     
SiO detections (filled circles) also appear in a wide range of color indices
in figure 4, but 
the SiO sources distribute with a steeper inclination ($\sim 1.98 \pm 0.35$) 
than the no-SiO subsample ($\sim 1.62 \pm 0.21$), 
indicating that the re-emission effect
at 12 $\mu$m is stronger in the SiO maser sources than the non-SiO emitting objects.  
The SiO maser is an indicator of active mass loss near the central star
($\sim 10^{14}$ cm),  suggesting the presence of thick dust re-emitting the
stellar radiation more at MIR wavelengths than the objects without SiO masers.

The above findings strongly suggest that these objects have 
an excessively large optical depth of circumstellar dust,
which is much larger than that of the circumstellar dust of OH 127.8+0.0. 
The order of magnitudes of the excess material can be estimated from $[3.6]-[12]$. 
We use the relation between $[3.6]-[12]$  and the mass loss rate calculated by 
\citet{kem02}. Applying the interstellar reddening corrections to $[3.6]-[12]$,
we computed the mass loss rate of the envelope from figure 6 of \citet{kem02},
and obtained the excess factor ($f_c$) of the circumstellar matter
(relative to that of the OH 127.8+0.0 envelope), 
which is listed in the last column of table 2. 
They exceed the optical depth of the OH 127.8+0.0 envelope by a factor of a few up to 80.
Uncertainty of the interstellar extinction correction of about 30 \% makes a 
shift of $[3.6]-[12]$ up to 0.7 at most, making the uncertainty of the excess factor 
to be within a factor of 2 even for the worst case.

The total envelope mass for each sampled object must exceed by a similar factor, i.e. 
these objects might have envelope masses of 0.4--20 $M_{\odot}$ if simply integrated until 
the outer radius of $\sim 10^{17}$ cm. The total mass of the envelope is proportional 
to the outer-boundary radius for the assumed density distribution 
of a constant-mass-loss model ($\rho \propto r^{-2}$), 
whereas the optical depth is rather sensitive to the inner-bounday radius of the dust envelope.
Because the color index, $[3.6]-[12]$, which was used here for evaluation of the excess factor,
does not give useful restrictions on the outer boundary radius in the Kemper et al's modeling 
(it is rather restricted by IRAS 60 or 100 $\mu$m flux density), 
the total mass of the thick matter estimated here has large ambiguity. 
Because, without precise modeling, it is uncertain as to how far from the central star
the extra material in the envelope is located, we only use the 
excess factor $f_c$ in this paper, which characterizes the envelope of these objects.


The total envelope mass for each sampled object can be reduced, in some degree,
by introducing a non-spherical distribution of the dust envelope, e.g. 
a torus or disk structure. However, even a wide opening angle of $120^{\circ}$ for cavity cones
can reduce the mass by 50\% compared 
with that of the spherical distribution with the same radial density profile.    
As discussed in the later sections,
high-velocity components of water maser emission may suggest the presence of the accretion disk
which creates collimated jets. In such a case, the scattered light 
through the pole of the torus should be observable, although
it is hard to detect the scattered light for these objects
because of the large interstellar extinction in the $K$ band.

\subsection{Characteristic of SiO maser emission}  It is striking that only the SiO $J=1$--0 $v=2$ line was
detected in 6 out of 8 SiO detected sources.  Even in the remaining two sources the intensity of the
$J=1$--0 $v=2$ line was considerably stronger than that of the  $J=1$--0 $v=1$ line. 
The trend, i.e. the increase of the $v=2$/1 line intensity ratio with the infrared color
[$C_{12}=log(F_{25}/F_{12})$, where $F_{12}$ and $F_{25}$ are IRAS 12 and 25 $\mu$m flux density],
was clearly first demonstrated by \citet{nak03b}. For the present sample of SiO detected sources, 
the MSX color, $C_{CE}=log(F_{21}/F_{12})$, is between 0.10 and 0.34, which fits well
with the above trend. All of the SiO detected sources were previously observed by the OH 1612 MHz line
and were all detected except for 18241$-$1010.    

Figure 4 indicates that, for the sampled objects,
SiO maser detection rates do not seem to correlate with the colors,
$[3.5]-[5.8]$, or $[3.6]-[12]$.
The flat detection rate seems to suggest that these stars are
still in a mass-losing stage, at the final transient stage of the AGB
to planetary-nebula phase. They are not at a later stage of the post-AGB phase 
when SiO masers should disappear. 

\subsection{New H$_2$O sources with wide spread emission}

Two H$_2$O maser sources, 18286$-$0959 and 18460$-$0151 (Fig.~3)
have a very wide velocity range ($\gtrsim$200 km s$^{-1}$) in their maser emission spectra.
Although the rich H$_2$O maser emission spectrum in IRAS 18286$-$0959 (Fig.~3 left) 
resembles the water maser spectra associated with compact H~{\sc ii} regions 
\citep[for example, ][]{kur05}, the nondetections of HCO$^+$, H53$\alpha$, and NH$_3$ lines 
toward these sources (the present work: table 3) suggest that 
they are not associated with dense molecular clouds (or compact H~{\sc ii} regions), 
and the rich H$_2$O emission could be the result of a very irregular velocity field in the envelope.
The fact that these sources are point-like with no associated nebulosity in their GLIMPSE images
suggests that they are stellar objects. 

Similar wide spread water maser spectra were found in the class of ``water fountain'' sources.   
There are four known water fountain sources [16342$-$3814, OH 12.8+0.9 (18139$-$1816), W43A (18450$-$0148),
19134+2131 \citep{lik88,bob05,ima04,ima05} with a few more probable cases \citep{gom94,dea06}. 
All of these objects exhibit the H$_2$O maser components in a velocity range that exceed 
the OH velocity range. The H$_2$O maser components are spatially more extended than
the OH maser components \citep{ima04,bob05}.  Among these ``water fountain sources'', 16342$-$3814 
had the highest velocity separation of $\sim 258$ km s$^{-1}$.  
Our object 18460$-$0151 definitely belongs to the class of water fountain sources from the similarity 
of the emission characteristic, and it has a record-high separation of 292  km s$^{-1}$ 
between emission components.

Both new objects, 18286$-$0959 and 18460$-$0151, are relatively bright IRAS sources (25 and 20 Jy at 12 $\mu$m respectively);
the former has a 2MASS counterpart, $J18312292-0957194$, with $K$ magnitude of 12.67 and $H-K=0.89$.
They have a rising mid-IR spectrum toward longer wavelengths, indicating the presence of a thick dust envelope.
From the radial velocities, we estimate kinematic distances of 18286$-$0959 and 18460$-$0151 as
3.1 and 6.8 kpc, respectively, and luminosities as $8.7 \times 10^3$ and $4.2 \times 10^4 \ L_{\odot}$ for these distances.

If these objects have a thick dust torus 
(as in the unified model of the Type I and II active galactic nuclei), 
optically observable objects \citep[as IRAS 16342$-$3814; ][]{sah99} 
must be seen from the polar-axis direction of the dust torus 
and unseen from equatorial directions.
If these high-velocity features are a part of the polar jet, 
which is created by the accretion disk, the higher velocity objects should have
bluer colors. 
This hypothesis seems applicable for 18286$-$0959 (because of the identified NIR counterpart
with $K=12.67$), but not for 18460$-$0151 (with no NIR counterpart).
Regarding 18460$-$0151, which has the highest velocity separation among the water fountains,
the large distance ($\sim 6.8$ kpc) and the strong interstellar extinction hide this object
behind molecular clouds. Applying the large interstellar extinction correction of $A_K=8.2$, 
we infer the extinction corrected $K$ magnitude to be fainter than 5.8.
It is slightly fainter than the extinction-corrected 3.6 $\mu$m magnitude ($[3.6]\sim 5.5$),
suggesting that the scattered light is not seen from this object.

\subsection{Envelopes of extremely red OH/IR stars}
The central stars of extremely red OH/IR objects
with low expansion velocity ($V_{exp}<15$ km s$^{-1}$) 
have been considered to be relatively low-initial-mass post-AGB stars 
compared with those of normal-color sources \citep{sev02}. 
In fact,  the red (``RI'') group sources of \citet{sev02} involved two objects in the present sample:
IRAS 18135$-$1456 and 18596+0315, which have the expansion velocities, 14.8 and 13.6 km s$^{-1}$, respectively.
Regarding expansion velocites derived from OH peaks, two water fountains, 18450$-$0148 (W43A) and 18460$-$0151,
are low expansion-velocity (lower-mass) sources ($V_{exp}=6$ and 11 km s$^{-1}$).

Figure 5 shows a plot of the excess factor of the envelope versus expansion velocity 
for OH 1612 MHz doubly-peaked objects in the present sample.  Note that
the excess factor was derived from $[3.6]-[12]$ (relative to OH 127.8+0.0; see section 3.1).  
It indicates that the excess factor does not correlate strongly with the expansion velocity 
of the envelope, and  hence the initial mass of the central star.

Figure 5 also indicates that objects with small expansion velocity ($V_{exp}<15$ km s$^{-1}$) seem 
to have lower SiO detection rates than the larger expansion-velocity objects ($V_{exp}>15$ km s$^{-1}$). 
This indicates that the lower-mass stars with small expansion velocities are at the phase
unfavorable for making strong SiO masers, e.g. dissociating SiO molecules by hot central star radiation,
and/or terminating mass loss from the central star.
This is consistent with the \citet{sev02}'s finding that
the red objects with small expansion velocites are stars in the early post-AGB phases 
with nearly zero mass loss; the extreme high-velocity flow develops at this phase and SiO masers 
gradually disappear. We infer that the red stars with large expansion velocites are 
AGB stars still in an active mass-losing stage because of the presence of SiO masers.
However both groups of stars, which are investigated in this paper,
are dressed by excessively large amounts of circumstellar matter irrespective 
of the mass of the central star.

The radii of the OH emitting regions were found to be 3 -- 16 $\times 10^{16}$ cm 
for some of the selected objects (OH 21.5+0.5, 30.1$-$0.7, and 32.0$-$0.5) 
with the OH phase-lag/angular-size measurements \citep{her85}. 
Because the H$_2$O masing region of water fountains is extended more than
the OH emission region \citep[for W43A]{ima02}, the thick material 
must be extended out of the OH masing region. 
There is evidence that H$_2$O maser outflow strikes the dense material
which is located outside of the OH masing region. 
Therefore, it is likely that the excess material found in section 3.1
is extended to considerably outer parts of the envelope, i.e. outside of the OH masing region.
The  H$_2$ number density of the OH masing region is known to be $\sim 10^3$--$10^4$ cm$^{-3}$ \citep[for example, ][]{net87},
and the density of the H$_2$O masing region must be, by several orders of magnitude, higher ($\sim 10^9$ cm$^{-3}$).
Thus, we cannot deny the possibility that the thick layer is extended
at the radius of more than $3\times 10^{16}$ cm, though
such a layer could be clumpy and dense, depending on the model. 
Envelope masses of some OH/IR stars [e.g., IRAS 18450$-$0148 (W43A)], 
exceed 4 $M_{\odot}$ even for the outer radius of $3\times 10^{16}$ cm. 
Though the envelope mass can vary by a factor of a few, 
depending on the uncertainty of interstellar extinction and 
nonspherical distribution of the thick layer as noted before, 
it is difficult to lower the envelope mass less than 1 $M_{\odot}$ 
for several of the thickest objects of the sample. 

A number of numerical modelings of thick circumstellar envelopes 
in the outgoing AGB phase have been made \citep[for example, ][]{van97,lor97}.
\citet{dav92} and \citet{suh97} examined superwind models to explain far IR spectra
of OH/IR objects accompanying deep silicate absorptions. One of their models, which has 
a dense shell between radii $2 \times 10^{16}$ and $4 \times 10^{16}$ cm 
in addition to a normal continuous flow \citep{suh97},  
successfully fits the computed to the observed spectral energy distribution of IRAS 18257$-$1000 (OH 21.5+0.5),
which has a relatively mild excess factor (1.4) in our list (table 2).
The large IRAS 60 $\mu$m flux densities of most of the other objects (though some of them 
might be contaminated from nearby clouds) 
entail for the much denser outer shell in these models, which inevitably results in a large total envelope mass
of more than a few $M_{\odot}$ \citep{dav92}; the superwind models artificially restrict the outer radius 
of the thick material to keep the enclosed mass smaller. 
Although the total mass of the outer envelope could in principle be observationally constrained
by CO $J=1$--0 intensities or IRAS 60/100 $\mu$m flux densities, it is hard
to estimate for these objects because of contamination by radiation from surrounding clouds.

Though all of these modelings of very red OH/IR objects assume a superwind 
with a mass loss rate of $\sim 10^{-4}\ M_{\odot}$ yr$^{-1}$ in a duration of about a few hundred years,
the origin of the thick material is not necessarily restricted to the superwinds which occur 
at the end of the AGB phase. Because no correlation appears between the excess factor and the expansion velocity,
the thick material does not seem to be related with the central star masses.
The excess material at the outer envelope of these objects can be a source of a shock front 
of the extremely high-velocity outflow created at the early post-AGB phase of stars, 
that is observed as water fountains. 
This may be due to gas ejection by binary-star interactions \citep{nor06}.
\citet{men02} found that the massive dust torus of the Red Rectangle  
has $M \sim 1.2\ M_{\odot}$, which was formed in the ejection of a common envelope after Roche lobe overflow. 
Similarly, stellar merging \citep[for example, ][]{bal05,tyl06} may create such massive circumstellar material.
Alternatively these stars may simply be in a dense dust cloud,  as found in IRAS 19312+1950 \citep{deg04a}.
We cannot deny such a possibility for some of these sources (but not all of them) 
because they are seen toward thick dust clouds. 
We infer here that massive circumstellar matter of the thickest OH/IR stars in the present sample 
can be formed by one of the aforementioned mechanisms.  

\section{Conclusion}
SiO maser and OH/IR objects with no 2MASS counterparts were identified 
in the GLIMPSE 3.6/4.5/5.8/8.0 $\mu$m images with a spatial resolution of 1.2\arcsec.  
Searches for SiO and H$_2$O masers lead to 10 new detections, each of which show  
characteristic maser spectra of the stronger SiO $J=1$--0 $v=2$ to $v=1$ line or
wide spread H$_2$O maser emission. It turned out that the dust envelopes of these objects 
are exceptionally thick, though they suffer from a large interstellar extinction.  
This suggests that a considerable mass of materials toward the objects 
must exist in the envelope. 
The SiO detection rate was uncorrelated with the mass of the envelope, but
the lower detection rate was obtained for the lower-expansion velocity subsample of 
the OH/IR stars. The latter suggests that the higher-expansion velocity objects
are AGB stars and lower-expansion velocity objects are more evloved stars such as post-AGB stars. 
The derived envelope mass is not correlated with this sequence of stellar evolution.       
From these observations, we believe that the thick material in some of these objects 
cannot be ascribed to the AGB-phase wind of the central star.

\acknowledgements
We thank Ed Churchwell and the GLIMPSE team for their help in the retrieval and processing of the survey data.  
We also thank the anonymous referee for useful comments for clarifing the content.
This research made use of the SIMBAD and VizieR databases operated at CDS, 
Strasbourg, France, and as well as use of data products from 
Two Micron All Sky Survey, which is a joint
project of the University of Massachusetts and Infrared Processing 
and Analysis Center/California Institute of Technology, 
funded by the National Aeronautics and Space Administration and
National Science foundation, and from the Midcourse Space 
Experiment at NASA/ IPAC Infrared Science Archive, which is operated by the 
Jet Propulsion Laboratory, California Institute of Technology, 
under contract with the National Aeronautics and Space 
Administration.   
This work is supported in part by a grant to SK from the Natural Sciences and Engineering Research Council of Canada.

\appendix
\section{IRAS 19027+0517}
The Glimpse images of the SiO maser source, IRAS 19027+0517,
exhibit conspicuous nebulosity around the central star at 5.8 and 8.0 $\mu$m (Fig.~6).
The nebula is lopsided to the central star and extended to the north. 
The surface brightness of the nebula seems to increase toward the central star, which suggests
a physical association of this nebula to the central star. 
Among 200 SiO sources investigated, only this object has an extended nebulosity.
The 2MASS images show a medium bright central star ($K=8.2$), but do not exhibit any extended features. 
The nebulous feature is considered to be due to the PAH emission with 
sharp spectral features at 6.2 and 7.8 $\mu$m \citep{ver01}. It is quite unusual that the O-rich
central star in the AGB phase (with SiO masers) accompanies such extended PAH nebulosity.
Therefore, we particularly investigated this object more with various molecular lines
with the NRO 45m telescope. In addition to the previous
detections of the SiO $J=1$--0  $v=1$ and 2 lines \citep{deg04b}, we found
emissions of the SiO $J=1$--0  $v=0$ and 3, and $J=2$--1 $v=1$ lines; the
spectra are shown in figure 7. Furthermore, we made 45$''$-grid 5-point mapping observations
by CO, HCN and HCO$^+$ $J=1$--0 lines, but no emission associated with the central star
was detected, except for the fore/background clouds with large velocity separations
(see figure 8). The observational results are summarized in table 5.   
Therefore, we conclude that most of the molecules are dissociated in the nebula toward IRAS 19027+0517,
an O-rich AGB star, by UV radiation from a closeby UV source creating the lopsided nebulosity.

\clearpage
\begin{deluxetable}{llllccccl}
\tabletypesize{\scriptsize}
\tablecolumns{9}
\tablewidth{0pc}
\tablecaption{Objects investigated.}
\tablehead{
Source  &   MSX6C name        & R.A.       &  Dec                  &   $F_{\rm C}$  &  Cac  &   Cce  &  SiO H$_2$O OH & OH name \\
        &                     & (h  m  s)  &  ($\circ$ $'$  $''$)  &                &       &        & (detection$^a$) & }
\startdata
18034$-$1858 & G011.0064$+$00.9220 &  18 06 25.3  & $-$18 57 44  &    8.8 &  0.194 &    0.186 &  n~   n~  --~ & \\
18100$-$1915 & G011.5218$-$00.5826 &  18 13 03.1  & $-$19 14 19  &   13.1 &  0.167 &    0.173 &  y~   n~  y$^3$ & OH11.52$-$0.58 \\
18135$-$1456 & G015.7005$+$00.7707 &  18 16 25.7  & $-$14 55 15  &   26.1 &  0.752 &    0.498 &y$^1$ y$^2$ y$^3$ & OH15.7+0.8  \\
18161$-$1713 & G013.9883$-$00.8613 &  18 19 01.7  & $-$17 12 07  &   25.3 &  0.283 &    0.083 &  n~  n~   n$^{10}$ & \\
18182$-$1504 & G016.1169$-$00.2903 &  18 21 07.0  & $-$15 03 22  &   48.5 &  0.190 &    0.268 &  y~ y$^4$ y$^3$ & OH16.1$-$0.3  \\
18198$-$1249 & G018.2955$+$00.4291 &  18 22 43.1  & $-$12 47 42  &   10.1 &  0.242 &    0.340 &  y~ n$^4$ y$^3$ & OH18.30+0.43\\
18212$-$1227 & G018.7683$+$00.3016 &  18 24 05.3  & $-$12 26 12  &    3.6 &  0.863 &    0.601 &  n~   n~  y$^3$ & OH18.8+0.3  \\
18241$-$1010 & G021.1164$+$00.7775 &  18 26 50.6  & $-$10 08 19  &   12.4 &  0.176 &    0.099 &  y~  --~  n$^{10}$ & \\
18245$-$1248 & G018.8384$-$00.5622 &  18 27 21.2  & $-$12 46 42  &    1.1 &  0.252 &    0.076 &y$^5$ -- -- & \\
18257$-$1000 & G021.4566$+$00.4911 &  18 28 31.0  & $-$09 58 15  &    9.6 &  0.172 &    0.305 &  y~ y$^{14}$ y$^3$ & OH021.457+00.491\\
18286$-$0959 & G021.7964$-$00.1273 &  18 31 22.9  & $-$09 57 20  &   45.0 &  0.181 & $-$0.129 &  n~   y~  y$^3$ & OH021.797$-$00.127\\
18298$-$0904 & G022.7482$+$00.0248 &  18 32 37.0  & $-$09 02 30  &   18.3 &  0.173 & $-$0.176 &  n~   n~  -- & \\
18325$-$0721 & G024.5814$+$00.2245 &  18 35 19.1  & $-$07 19 23  &    4.7 &  0.178 &    0.145 &  n~   n~  y$^3$ & OH24.6+0.2\\
18327$-$0645 & G025.1301$+$00.4841 &  18 35 24.3  & $-$06 42 59  &   10.1 &  0.349 & $-$0.099 &  n~   n~  --~ & \\
18407$-$0358 & G028.5203$-$00.0141 &  18 43 25.8  & $-$03 55 55  &    3.8 &  0.225 &    0.197 &  y~   n~  y$^3$ & OH28.5$-$0.0\\
18424$-$0509 & G027.6621$-$00.9179 &  18 45 04.9  & $-$05 06 28  &   14.5 &  0.111 & $-$0.146 &  n~   n~  --~ & \\
18444$-$0359 & G028.9304$-$00.8287 &  18 47 05.2  & $-$03 56 21  &    7.8 &  0.166 & $-$0.135 &  n~   n~  --~ & \\
18450$-$0148 & G030.9441$+$00.0350 &  18 47 41.2  & $-$01 45 11  &   23.9 &  0.978 &    0.519 &y$^1$ y$^8$ y$^3$ & OH31.8+0.0, W43A \\
18460$-$0151 & G031.0126$-$00.2195 &  18 48 43.0  & $-$01 48 30  &   14.7 &  0.205 &    0.176 &  n~   y~  y$^3$ & OH30.1$-$0.2 \\
18460$-$0254 & G030.0908$-$00.6866 &  18 48 42.0  & $-$02 50 29  &  127.9 &  0.184 &    0.261 &y$^7$ y$^{14}$ y$^3$ & OH30.1$-$0.7 \\
18488$-$0107 & G031.9844$-$00.4849 &  18 51 26.2  & $-$01 03 52  &   31.1 &  0.177 &    0.196 &y$^1$ y$^{14}$ y$^3$ & OH32.0$-$0.5 \\
18509$-$0018 & G032.9524$-$00.5687 &  18 53 30.0  & $-$00 14 28  &   24.9 &  0.211 &    0.122 &y$^1$  --~ y$^3$ & OH32.95$-$0.57\\
18517$+$0037 & G033.8728$-$00.3350 &  18 54 20.8  & $+$00 41 05  &   25.9 &  0.244 &    0.187 &y$^1$ y$^9$ y$^3$ & OH033.873$-$00.335\\
18596$+$0315 & G037.1185$-$00.8473 &  19 02 06.3  & $+$03 20 16  &    2.9 &  0.784 &    0.501 &  n~   n~  y$^3$ & OH37.1$-$0.8 \\
19006$+$0624 & G040.0220$+$00.3818 &  19 03 03.4  & $+$06 28 54  &    2.2 &  0.276 &    0.060 &y$^1$  y$^{14}$ y$^{10}$ & OH40.02+0.38\\
19011$+$0638 & G040.2793$+$00.3766 &  19 03 33.1  & $+$06 42 29  &   11.4 &  0.297 &    0.206 &  n~   n~  --~ & \\
19025$+$0702 & G040.8005$+$00.2455 &  19 04 59.0  & $+$07 06 40  &    6.9 &  0.354 &    0.068 &  n~  --~  --~ & \\
19087$+$1006 & G044.2404$+$00.3090 &  19 11 10.0  & $+$10 11 37  &    3.0 &  0.594 &    0.259 &  n~   n~  n$^{12}$ & \\
19112$+$1220 & G046.4992$+$00.8092 &  19 13 37.4  & $+$12 25 39  &    5.2 &  0.383 &    0.058 &  n~   n~  n$^{10}$ & \\
19114$+$0920 & G043.8675$-$00.6247 &  19 13 49.4  & $+$09 25 51  &   10.5 &  0.157 & $-$0.172 &  n~   n~  n$^{10}$ & \\
19180$+$1230 & G047.4257$-$00.5624 &  19 20 22.0  & $+$12 36 24  &    6.1 &  0.231 & $-$0.123 &  n~  --~  --~ & \\
19254$+$1631 & G051.8042$-$00.2247 &  19 27 42.1  & $+$16 37 25  &   22.1 &  0.226 &    0.276 &y$^1$ --~  y$^{13}$ & OH51.8$-$0.2\\
19254$+$1724 & G052.5814$+$00.2014 &  19 27 41.1  & $+$17 30 36  &    3.4 &  0.675 &    0.273 &  n~  --~  n$^{10}$ & \\
19266$+$1815 & G053.4614$+$00.3547 &  19 28 53.1  & $+$18 21 23  &    8.8 &  0.168 & $-$0.168 &  n~   --~ --~ & \\
19275$+$1702 & G052.5042$-$00.4085 &  19 29 46.5  & $+$17 09 01  &    9.3 &  0.185 & $-$0.136 &  n~   n~  --~ & \\
19440$+$2251 & G059.4784$-$00.8969 &  19 46 09.2  & $+$22 59 24  &   17.4 &  0.153 &    0.258 &y$^1$ n$^8$ y$^{11}$ & OH59.48$-$0.90\\
G014.2758    & G014.2758$-$00.1602 &  18 17 01.0  & $-$16 37 00  &    6.1 &  0.183 &    0.122 &  n~   n~  --~ & \\
G017.3913    & G017.3913$-$00.2891 &  18 23 35.0  & $-$13 55 49  &    5.7 &  0.080 &    0.240 &  y~ n$^6$ y$^3$ & OH17.4$-$0.3\\
G022.0425    & G022.0425$-$00.6084 &  18 33 34.6  & $-$09 57 36  &    1.2 &  0.628 &    0.632 &  n~ y$^2$ y$^3$ & OH22.1$-$0.6\\
G024.6610    & G024.6610$+$00.0868 &  18 35 57.5  & $-$07 18 58  &    3.1 &  0.218 &    0.324 &  y~ n$^6$ y$^3$ & OH24.7+0.1\\
G027.8433    & G027.8433$-$00.1060 &  18 42 30.9  & $-$04 34 35  &    7.1 &  0.081 & $-$0.207 &  n~   n~  --~ & \\
\enddata
\tablenotetext{a}{"y" or "n" indicates the detection or nondetection in this paper, othewise noted.
References:
$^1$: \citet{nak03b}, $^2$: \citet{eng86}, $^3$: \citet{sev01},
$^4$: \citet{gom90}, $^5$: \citet{izu99}, $^6$: \citet{nym86}, 
$^7$: \citet{nym98}, $^8$: \citet{lik92}, $^9$: \citet{eng96},
$^{10}$: \citet{tel91},$^{11}$: \citet{ede88}, $^{12}$: \citet{lew87},
$^{13}$: \citet{che93}, $^{14}$: \citet{dea06}.} 
\end{deluxetable}


\begin{deluxetable}{lccrrrrrrrc}
\tabletypesize{\scriptsize}
\tablecolumns{11}
\tablewidth{0pc}
\tablecaption{Objects identified with Spitzer-Glimpse Survey.}
\tablehead{
Source             &   $l$    & $b$   & $F_{3.6}$  & $F_{4.5}$ & $F_{5.8}$   & $F_{8.0}$  & [3.6]  & [3.6] & $A_K$ & $f_{c}$  \\
(IRAS or MSX)      &    ($\circ$) & ($\circ$) &  (mJy)  &  (mJy) &   (mJy)   &  (mJy)   & $-[5.8]$ & $-[12]$ &    &   }
\startdata
18034$-$1858  & +11.0061  & $+$0.9222  &    29.9  &  308.8 &   1617.0 &  ---     &5.30 &  8.74 &  1.50  &  7.6\\ 
18100$-$1915  & +11.5216  & $-$0.5824  &    72.7  &  712.7 &   3641.0 &  ---     &5.21 &  8.21 &  3.05  &  4.9\\ 
18135$-$1456  & +15.7011  & $+$0.7706  &    26.2  &  166.6 &   1033.0 &  ---     &4.95 & 10.06 &  1.74  & 15.9\\ 
18161$-$1713  & +13.9884  & $-$0.8611  &   264.8  & 1988.0 &  11510.0 &  ---     &5.06 &  7.52 &  3.33  &  3.2\\ 
18182$-$1504  & +16.1173  & $-$0.2907  &   512.4  & ---    &  14120.0 &  ---     &4.57 &  7.51 &  5.64  &  2.6\\ 
18198$-$1249  & +18.2956  & $+$0.4291  &    56.2  &  847.5 &   1818.0 &  ---     &4.74 &  8.20 &  2.49  &  5.1\\ 
18212$-$1227  & +18.7688  & $+$0.3017  &     1.9  &    9.9 &     77.7 &   562.0  &4.99 & 10.76 &  3.20  & 20.9\\ 
18241$-$1010  & +21.1166  & $+$0.7775  &   676.4  & 2918.0 &   9559.0 &  ---     &3.84 &  5.73 &  1.65  &  1.3\\ 
18245$-$1248  & +18.8387  & $-$0.5622  &    81.2  &  447.1 &   1468.0 &  1750.0  &4.11 &  5.40 &  4.91  &  0.8\\ 
18257$-$1000  & +21.4565  & $+$0.4911  &   434.1  & ---    &   ---    &  ---     &     &  5.93 &  2.63  &  1.4\\ 
18286$-$0959  & +21.7972  & $-$0.1272  &   541.6  & ---    &   ---    &  ---     &     &  7.37 &  7.71  &  2.0\\ 
18298$-$0904  & +22.7483  & $+$0.0252  &   231.0  & ---    &   3409.0 &  ---     &3.89 &  7.31 &  7.15  &  2.1\\ 
18325$-$0721  & +24.5814  & $+$0.2243  &   198.1  &  600.7 &   2845.0 &  ---     &3.86 &  6.01 & 11.48  &  0.3\\ 
18327$-$0645  & +25.1301  & $+$0.4842  &    42.9  &  270.2 &   1074.0 &  ---     &4.46 &  8.50 &  4.09  &  5.3\\ 
18407$-$0358  & +28.5203  & $-$0.0143  &    75.3  & ---    &   2288.0 &  ---     &4.67 &  6.82 & 10.74  &  1.1\\ 
18424$-$0509  & +27.6622  & $-$0.9174  &    98.3  &  463.1 &   1573.0 &  ---     &3.98 &  7.99 &  1.31  &  5.0\\ 
18444$-$0359  & +28.9303  & $-$0.8288  &   232.9  & ---    &   3738.0 &  ---     &3.98 &  6.38 &  1.77  &  1.9\\ 
18450$-$0148  & +30.9439  & $+$0.0351  &     1.0  &   12.5 &   0217.2 &  2564.0  &6.81 & 13.51 & 14.79  & 37.8\\ 
18460$-$0151  & +31.0124  & $-$0.2194  &    23.9  &  374.8 &   2598.0 &  ---     &6.06 &  9.54 &  8.23  &  6.7\\ 
18460$-$0254  & +30.0910  & $-$0.6865  &   558.9  & 5860.0 &   ---    &  ---     &     &  8.47 &  3.35  &  5.5\\ 
18488$-$0107  & +31.9845  & $-$0.4853  &   778.6  & 6270.0 &  18690.0 &  ---     &4.42 &  6.57 &  2.70  &  2.0\\ 
18509$-$0018  & +32.9528  & $-$0.5689  &   273.0  & 1198.0 &   3568.0 &  3952.0  &3.76 &  7.47 &  1.59  &  3.6\\ 
18517$+$0037  & +33.8727  & $-$0.3353  &   107.9  & ---    &   4579.0 &  ---     &5.03 &  8.52 &  3.62  &  5.6\\ 
18596$+$0315  & +37.1184  & $-$0.8474  &     9.9  &   49.7 &    217.3 &   806.0  &4.32 &  8.73 &  1.54  &  7.5\\ 
19006$+$0624  & +40.0224  & $+$0.3813  &    85.7  &  484.3 &   1566.0 &  ---     &4.12 &  6.09 &  2.76  &  1.5\\ 
19011$+$0638  & +40.2794  & $+$0.3761  &    84.9  &  939.7 &   5077.0 &  ---     &5.41 &  7.89 &  2.71  &  4.2\\ 
19025$+$0702  & +40.8006  & $+$0.2452  &    22.1  &  187.6 &   1009.0 &  4079.0  &5.11 &  8.80 &  2.54  &  7.2\\ 
19087$+$1006  & +44.2406  & $+$0.3086  &    00.9  &   14.7 &    146.5 &   971.8  &6.49 & 11.37 &  3.80  & 28.3\\ 
19112$+$1220  & +46.4995  & $+$0.8093  &    01.6  &   23.5 &    193.8 &  1029.0  &6.17 & 11.35 &  1.88  & 32.9\\ 
19114$+$0920  & +43.8677  & $-$0.6248  &   192.3  &  865.6 &   2582.0 &  1809.0  &3.78 &  6.91 &  2.76  &  2.4\\ 
19180$+$1230  & +47.4258  & $-$0.5625  &    93.5  &  513.5 &   1714.0 &  ---     &4.12 &  7.10 &  2.92  &  2.6\\ 
19254$+$1631  & +51.8039  & $-$0.2249  &   155.1  & 1076.0 &   4806.0 &  ---     &4.69 &  7.95 &  2.80  &  4.3\\ 
19254$+$1724  & +52.5815  & $+$0.2014  &     0.2  &   02.4 &     44.4 &   503.2  &6.83 & 13.14 &  2.40  & 88.4\\ 
19266$+$1815  & +53.4611  & $+$0.3551  &    42.8  &  253.3 &    910.0 &   606.5  &4.28 &  8.35 &  2.03  &  5.8\\ 
19275$+$1702  & +52.5040  & $-$0.4085  &   359.6  & 1483.0 &   4437.0 &  ---     &3.69 &  6.10 &  2.85  &  1.5\\ 
19440$+$2251  & +59.4784  & $-$0.8966  &   179.4  & 1460.0 &   ---    &  ---     &     &  7.53 &  0.94  &  4.0\\ 
G014.2758     & +14.2760  & $-$0.1600  &   367.1  & ---    &   ---    &  ---     &     &  5.62 &  8.35  &  0.4\\ 
G017.3913     & +17.3915  & $-$0.2892  &    83.4  &  583.4 &   3113.0 &  ---     &4.89 &  7.15 &  5.82  &  2.1\\ 
G022.0425     & +22.0428  & $-$0.6085  &     6.6  &   22.0 &     83.5 &   365.7  &3.72 &  8.22 &  3.91  &  4.6\\ 
G024.6610     & +24.6608  & $+$0.0867  &     6.1  &  130.0 &    990.5 &  ---     &6.49 &  9.33 & 11.05  &  4.7\\ 
G027.8433     & +27.8430  & $-$0.1061  &   236.1  &  811.8 &   2035.0 &  2720.0  &3.30 &  6.26 &  6.30  &  1.2\\ 
\enddata
\end{deluxetable} 

\renewcommand{\arraystretch}{0.7}
\begin{table*}
\caption{Detections by the SiO or H$_2$O maser line.}\label{tab:table3}
\begin{tabular}{llrrrr}
\tableline\tableline
Source   & Transition  & $V_{lsr}$ & $Ta^{*}$  & line flux  & rms   \\
                    &  & (km s$^{-1}$) & (K) &  (K km s$^{-1}$)  &  (K) \\
\tableline   
18100$-$1915 & SiO $J=1$--0 $v=1$  &  ---  &  ---   &  ---   & 0.121 \\                 
             & SiO $J=1$--0 $v=2$  &  16.5 &  1.202 &  2.087 & 0.114 \\
             & H$_2$O $6_{16}$--$5_{23}$  &   --- &    --- &   ---  & 0.109 \\
18182$-$1504 & SiO $J=1$--0 $v=0$  &  18.3 &  0.650 &  2.536 & 0.153 \\
             & SiO $J=1$--0 $v=1$  &  22.9 & 16.661 & 45.758 & 0.166 \\
             & SiO $J=1$--0 $v=2$  &  22.5 & 41.500   & 130.23 & 0.168 \\
             & SiO $J=1$--0 $v=3$  &  22.5 & 10.434 & 28.839 & 0.173 \\
             & $^{29}$SiO $J=1$--0 $v=1$  &  23.6 &  1.093 &  2.623 & 0.151 \\
18198$-$1249 & SiO $J=1$--0 $v=1$  &  ---  &  ---   &  ---   & 0.104 \\
             & SiO $J=1$--0 $v=2$  &  48.2 &  0.579 & 0.429  & 0.107 \\
18241$-$1010 & SiO $J=1$--0 $v=1$  &  ---  &  ---   & ---    & 0.084 \\ 
             & SiO $J=1$--0 $v=2$  & 116.3 &  0.620 & 2.795  & 0.085 \\
18257$-$1000 & SiO $J=1$--0 $v=1$  &  ---  &  ---   & ---    & 0.097 \\
             & SiO $J=1$--0 $v=2$  & 114.9 &  0.781 &  1.427 & 0.097 \\  
18286$-$0959 & SiO $J=1$--0 $v=1$  &  ---  &  ---   & ---    & 0.123 \\ 
             & SiO $J=1$--0 $v=2$  & --- &  --- &  --- & 0.118 \\   
             & H$_2$O $6_{16}$--$5_{23}$  & $-14.7$ & 3.863 & 166.3 & 0.043 \\
             & SiO $J=2$--1 $v=1$ &  ---  &  ---   & ---    & 0.065 \\ 
             & HCO$^+$ $J=1$--0    &  ---  &  ---   & ---    & 0.068 \\
18407$-$0358 & SiO $J=1$--0 $v=1$  & 106.5 &  0.572 &  1.457 & 0.108 \\
             & SiO $J=1$--0 $v=2$  & 107.3 &  0.921 &  2.126 & 0.097 \\
             & H$_2$O $6_{16}$--$5_{23}$  &   --- &    --- &   ---  & 0.087 \\
18460$-$0151 & SiO $J=1$--0 $v=1$  & ---   &  ---   &  ---   & 0.100 \\  
             & SiO $J=1$--0 $v=2$  & ---   &  ---   &  ---   & 0.084 \\  
             & H$_2$O $6_{16}$--$5_{23}$  & 117.0  &  4.881  & 24.38  & 0.050 \\ 
             & SiO $J=2$--1 $v=1$ &  ---  &  ---   & ---    & 0.069 \\ 
             & HCO$^+$ $J=1$--0    &  ---  &  ---   & ---    & 0.084 \\
G017.3913    & SiO $J=1$--0 $v=1$  &  ---  &  ---   & ---    & 0.119 \\
             & SiO $J=1$--0 $v=2$  &  28.7 &  1.398 & 4.106  & 0.117 \\
G024.6610    & SiO $J=1$--0 $v=1$  & ---   &  ---   &  ---   & 0.103 \\
             & SiO $J=1$--0 $v=2$  &  57.7 &  1.247 &  3.662 & 0.101 \\
\tableline  
\end{tabular}  
\end{table*}
\newpage
\begin{table*}
\caption{Negative results for the SiO and H$_2$O maser line search.}\label{tab:table4}
  \begin{tabular}{lccc}
  \tableline\tableline
Source  &  rms    & rms & rms\tablenotemark{a}  \\
        & (SiO $v=1$) &  (SiO $v=2$) & (H$_2$O) \\
        &   (K)          & (K)     & (K) \\
\tableline   
18034$-$1858  &  0.115  &   0.114  & 0.096 \\
18161$-$1713  &  0.119  &   0.115  & 0.107 \\
18212$-$1227  &  0.096  &   0.095  & 0.086 \\
18298$-$0904  &  0.097  &   0.105  & 0.075 \\
18325$-$0721  &  0.075  &   0.073  & 0.087 \\
18327$-$0645  &  0.099  &   0.097  & 0.088 \\
18424$-$0509  &  0.357  &   0.396  & 0.088 \\ 
18444$-$0359  &  0.126  &   0.121  & 0.113 \\
18596$+$0315  &  0.117  &   0.115  & 0.086 \\  
19011$+$0638  &  0.113  &   0.102  & 0.084 \\
19025$+$0702  &  0.111  &   0.113  & --- \\
19087$+$1006  &  0.077  &   0.075  & 0.080 \\
19112$+$1220  &  0.088  &   0.084  & 0.082 \\
19114$+$0920  &  0.086  &   0.086  & 0.151 \\
19180$+$1230  &  0.071  &   0.073  & --- \\
19254$+$1724  &  0.094  &   0.082  & --- \\ 
19266$+$1815  &  0.077  &   0.071  & --- \\ 
19275$+$1702  &  0.094  &   0.090  & 0.093 \\
G014.2758     &  0.109  &   0.109  & 0.099 \\
G022.0425     &  0.075  &   0.078  & --- \\
G027.8433     &  0.103  &   0.099  & 0.062 \\
\tableline 
\end{tabular}  
\tablenotetext{a}{Dash indicates not observed.}
\end{table*}

\clearpage
\begin{table*}
\caption{Observation summary for IRAS 19027+0517.}\label{tab:table5}
\begin{tabular}{lccccc}
\tableline\tableline
Transition & rest freq.   & $V_{lsr}$   & $T_a^*$  & line flux  & rms     \\
           &   (GHz)      & (km s$^{-1}$) & (K) &  (km s$^{-1}$ K)  &  (K) \\
\tableline   
H$_2$O $6_{16}$--$5_{23}$& 22.23508 & ---   &   ---  &    --- & 0.073 \\
NH$_3$ $1_1$--$1_1$              & 23.694   &   --- &    --- &    --- & 0.063 \\
NH$_3$ $2_2$--$2_2$              & 23.722   &   --- &    --- &    --- & 0.044 \\
NH$_3$ $3_3$--$3_3$              & 23.870   &   --- &    --- &    --- & 0.052 \\
SiO $J=1$--0 $v=0$       & 43.42386 &  34.2 & 0.173  &  0.417 & 0.053 \\
SiO $J=1$--0 $v=1$       & 43.12208 &  31.7 & 1.288  &  3.348 & 0.058 \\
SiO $J=1$--0 $v=2$       & 42.82059 &  31.7 & 1.202  &  3.826 & 0.067 \\
SiO $J=1$--0 $v=3$       & 42.51938 &  31.6 & 0.298  &  0.557 & 0.053 \\
$^{29}$SiO $J=1$--0 $v=1$& 42.87992 & ---   &  ---   &  ---   & 0.054 \\
H53$\alpha$              & 42.95197 & ---   &  ---   &  ---   & 0.062 \\
SiO $J=2$--1 $v=1$       & 86.24342 & 32.3  & 0.204  & 0.584  & 0.034 \\
SiO $J=2$--1 $v=0$       & 86.84700 & ---   &  ---   &  ---   & 0.037 \\
HCN $J=1$--0             & 88.63185 &   ---\tablenotemark{a} &    --- &    --- & 0.053 \\  
HCO$^+$ $J=1$--0         & 89.18852 &   --- &    --- &    --- & 0.061 \\ 
CO  $J=1$--0             &115.27120 &   ---\tablenotemark{a} &    --- &    --- & 0.200 \\
\tableline
\end{tabular}  
\tablenotetext{a}{Contaminations at $V_{lsr}=12$, 46, and 63 km s$^{-1}$.}
\end{table*}

\clearpage
\begin{figure*}
\epsscale{1.5}
\plotone{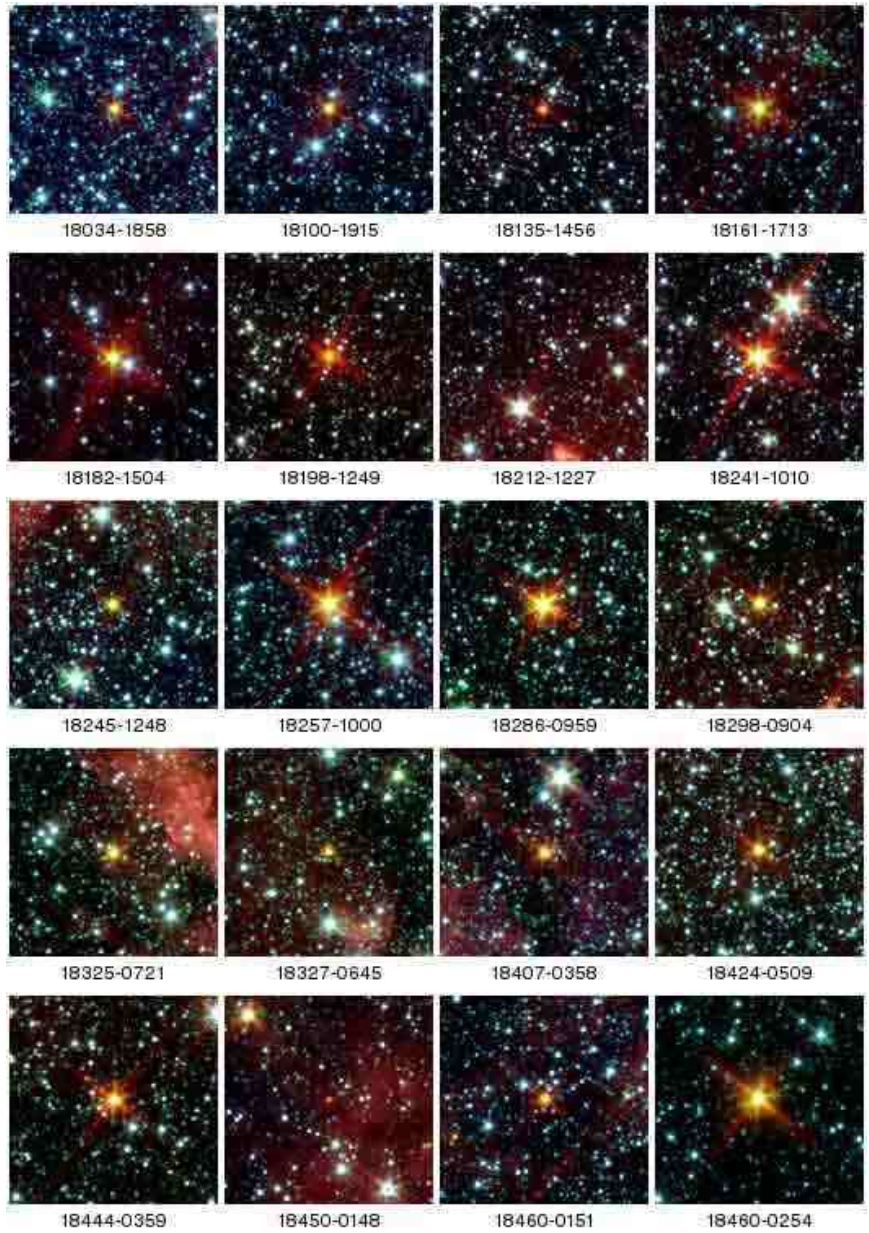}
  \caption{a. Composite-color images ($285''\times 285''$) of the GLIMPSE field of the sources in Table 1. 
IRAC 3.6, 4.5 and 5.8 $\mu$m bands are represented by blue, green, and red colors respectively.
  Objects are at the center of the images, and the directions of increasing Galactic longitude and latitude 
  are left and up, respectively.
  \label{fig:GLIMPASEa}}
\end{figure*}
\clearpage
\setcounter{figure}{0}
\begin{figure*}
\epsscale{1.5}
\begin{center}
\plotone{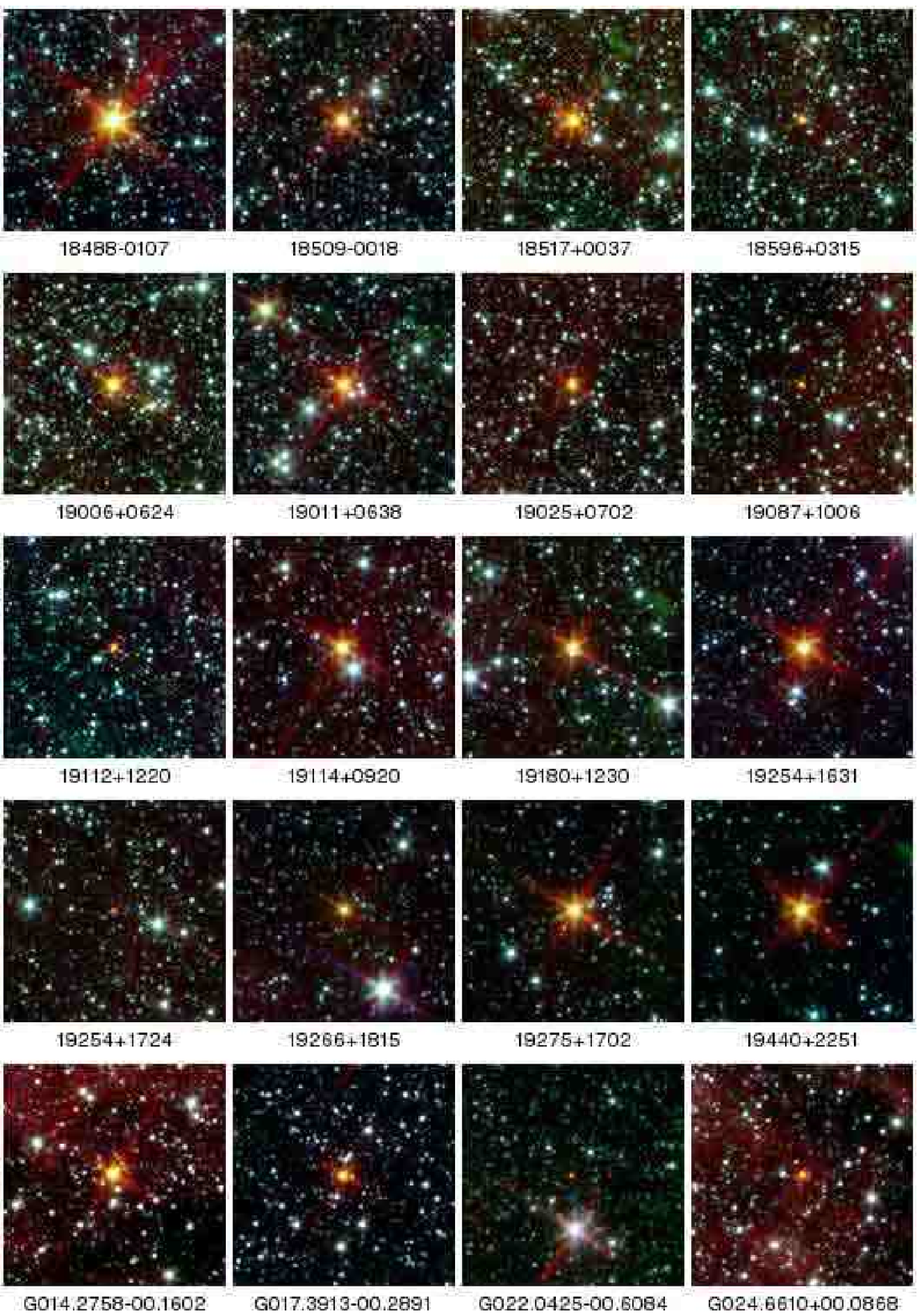}\\[5mm]
\vspace{1cm}
\epsscale{0.36}
\plotone{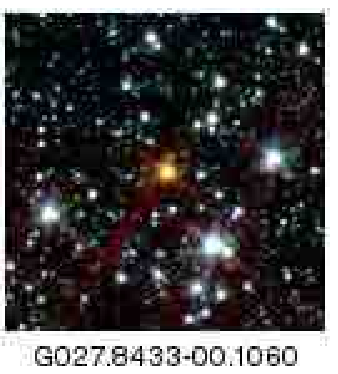}\\[5mm]
\caption{b. and c. --- Continued.\label{fig:GLIMPASEbc}}
\end{center}
\end{figure*}
\begin{figure*}
\epsscale{1.6}
\plotone{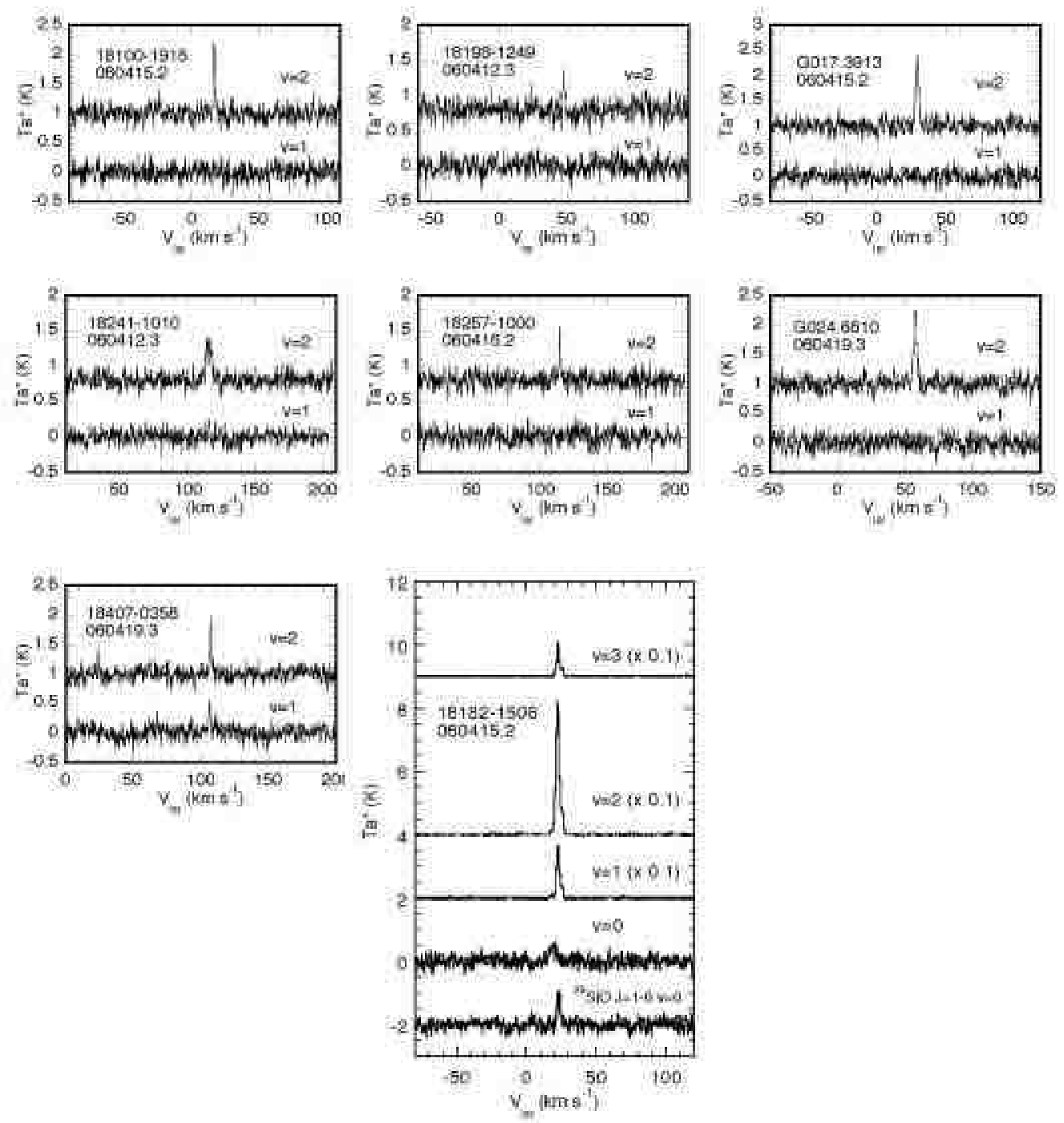}
  \caption{SiO $J=1$--0 $v=1$ and 2 spectra of the detected sources.
  For IRAS 18182$-$1504 the SiO $J=1$--0 $v=0$ 1,  2, and 3, 
  and $^{29}$SiO $J=1$--0 $v=0$ spectra are shown. Source name and observed date
  (in yymmdd.d format) are indicated on the left of each panel.
  were indicated
  }\label{fig:SiO}
\end{figure*}

\begin{figure*}
\epsscale{1.4}
\plotone{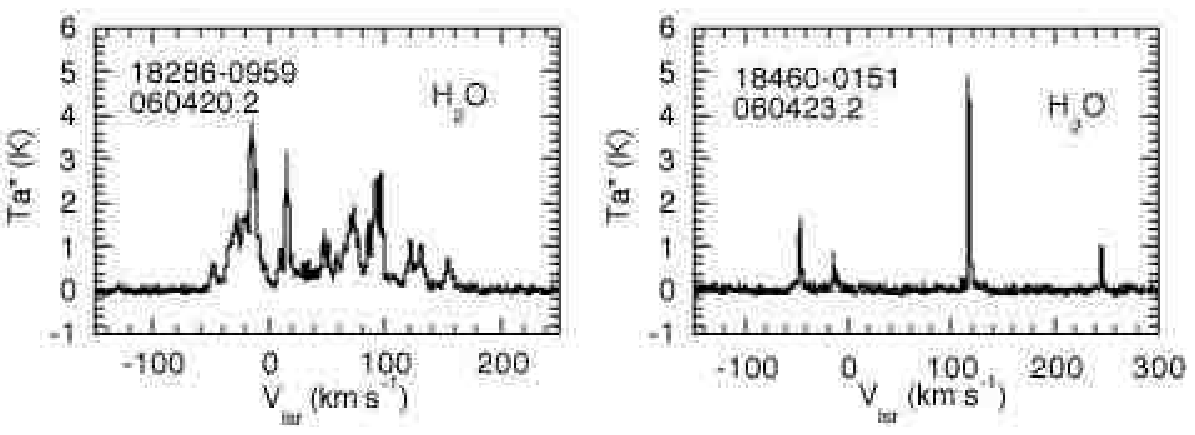}
  \caption{H$_2$O $6_{16}$--$5_{23}$ spectra for the detected sources. 
  Source name and observed date (in yymmdd.d format) are indicated on the left of each panel.
  }\label{fig:H2O}
\end{figure*}

\begin{figure*}
\epsscale{0.7}
\plotone{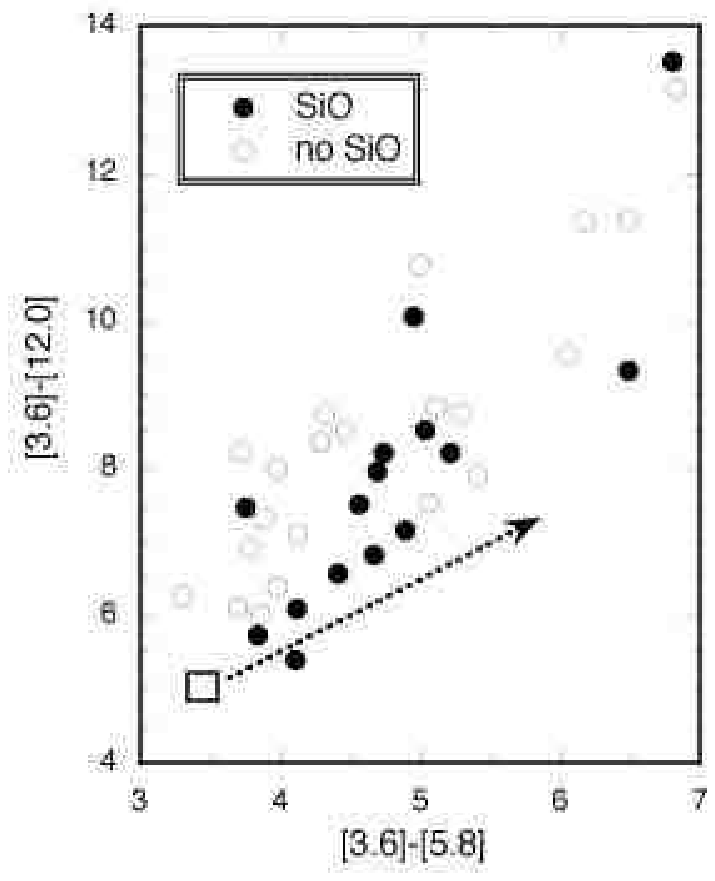}
  \caption{Two-color diagram, $[3.6]-[12]$ versus $[3.6]-[5.8]$.
The filled and unfilled circles indicate SiO detection and nondetection,
and the square indicates the position of OH 127.8+0.0. The dotted arrow indicates
direction of interstellar reddening; the length corresponds to $A_K=15$.
  }\label{fig:two-color}
\end{figure*}
\begin{figure*}
\epsscale{0.7}
\plotone{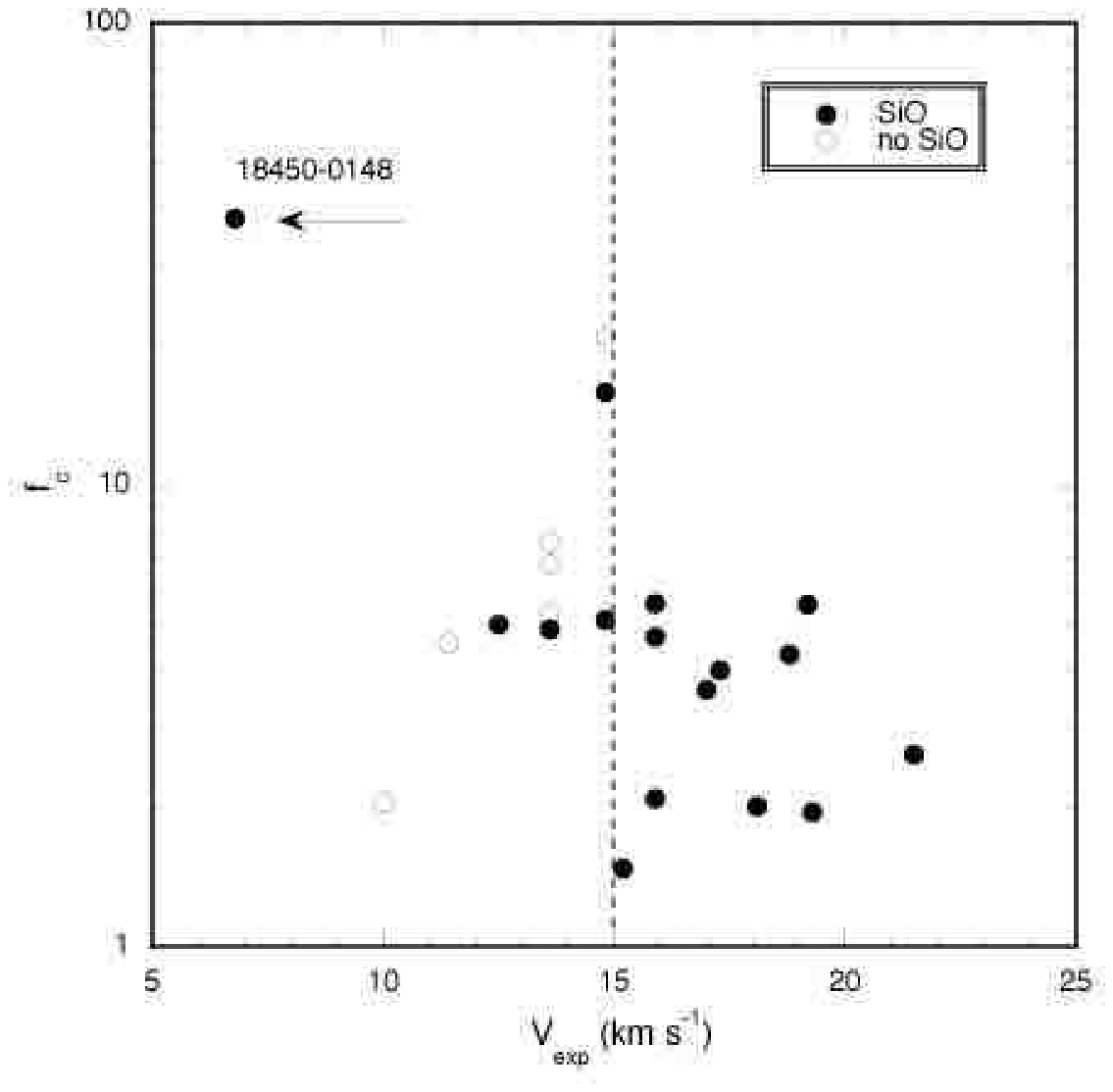}
  \caption{Excess factor versus expansion velocity of the OH 1612 MHz double-peak objects. 
  Filled and unfilled circles indicate SiO detection and nondetection. The broken line
  indicates a separation for the low-mass stars. 
  }\label{fig: mass_loss_Vepxp}
\end{figure*}

\begin{figure*}
\epsscale{0.5}
\plotone{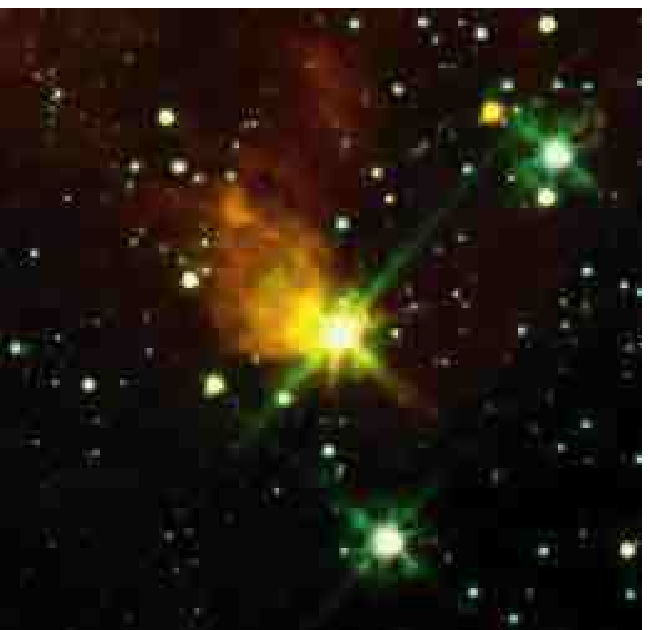}
  \caption{Glimpse color-composite image ($285''\times 285''$) of IRAS 19027+0517 made from 4.5 (blue), 5.8(green), and 8.0 (red) $\mu$m IRAS image.
  The directions of increasing Galactic longitude and latitude are left and up, respectively.
  }\label{fig:19207+0517image}
\end{figure*}
\begin{figure*}
\epsscale{1.2}
\plotone{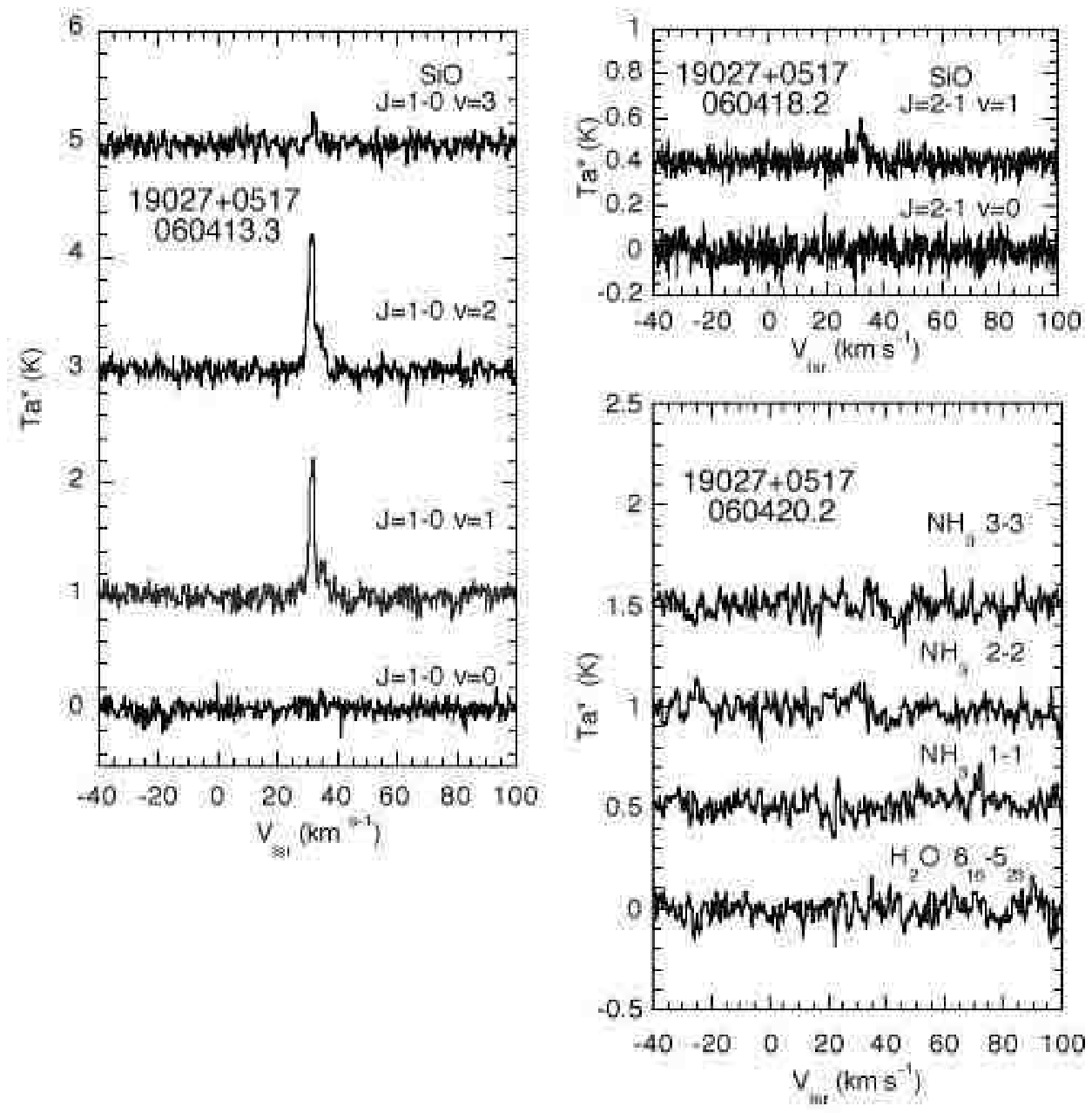}
  \caption{SiO, H$_2$O and NH$_3$ spectra (a)
  CO $J=1$--0 spectra (b) toward IRAS 19027+0517. The number between the parenthesis
  on the left indicates R.A. and Dec. offsets in arcsec from the star position.
  The top  is the spectrum of the bottom (center) subtracted 
  by the average of surrounding 4 positions (middle).
  }\label{fig:19207+0517SiOH2O}
\end{figure*}
\begin{figure*}
\epsscale{0.6}
\plotone{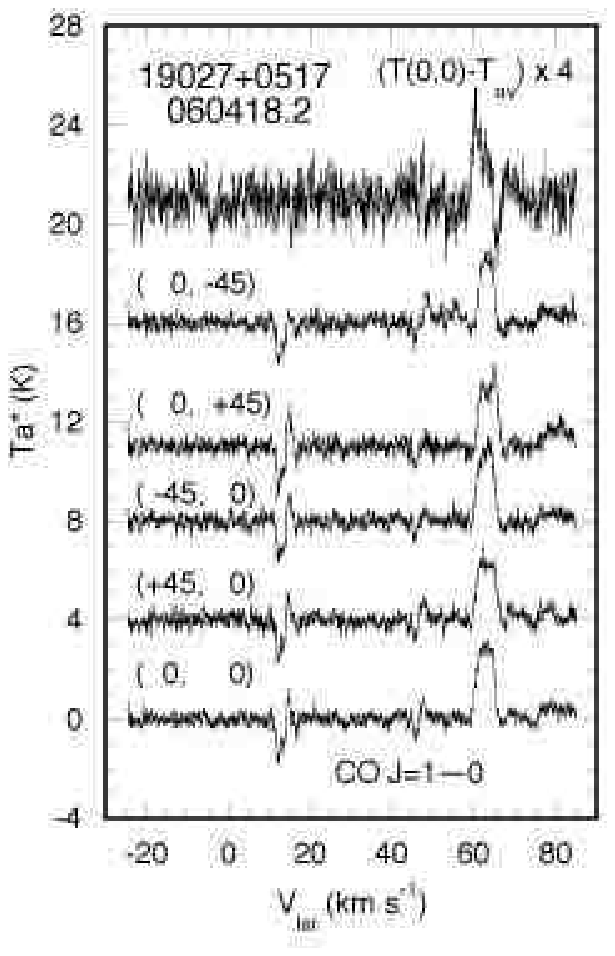}
  \caption{CO $J=1$--0 spectra toward IRAS 19027+0517. The number between the parenthesis
  on the left indicates R.A. and Dec. offsets in arcsec from the star position.
  The top  is the spectrum of the bottom (center) subtracted 
  by the average of surrounding 4 positions (middle).
  }\label{fig:19207+0517CO}
\end{figure*}

\end{document}